\documentclass[sigconf]{acmart}

\usepackage{booktabs} 

\usepackage[utf8]{inputenc}

\usepackage{caption}
\usepackage{subcaption}
\usepackage{tabularx}

\setcopyright{rightsretained}

\usepackage{flushend}

\acmConference[KDD Workshop on AI for Fashion]{KDD}{2018}{London, UK}

\usepackage{soul}
\usepackage{color}

\begin{document}
\title{A Hierarchical Deep Learning Natural Language Parser for Fashion}

\author{José Marcelino}
\affiliation{%
 \institution{Farfetch}
 \institution{Universidade de Coimbra}
}
\email{jose.marcelino@farfetch.com}

\author{João Faria}
\affiliation{%
 \institution{Farfetch}
}
\email{joao.faria@farfetch.com}

\author{Luís Baía}
\affiliation{%
 \institution{Farfetch}
}
\email{luis.baia@farfetch.com}

\author{Ricardo Gamelas Sousa}
\orcid{0000-0001-8822-5412}
\affiliation{%
 \institution{Farfetch}
}
\email{ricardo.sousa@farfetch.com}

\renewcommand{\shortauthors}{J. Marcelino et al.}

\begin{abstract}
This work presents a hierarchical deep learning natural language parser for fashion. Our proposal intends not only to recognize fashion-domain entities but also to expose syntactic and morphologic insights. We leverage the usage of an architecture of specialist models, each one for a different task (from parsing to entity recognition). Such architecture renders a hierarchical model able to capture the nuances of the fashion language. The natural language parser is able to deal with textual ambiguities which are left unresolved by our currently existing solution.
Our empirical results establish a robust baseline, which justifies the use of hierarchical architectures of deep learning models while opening new research avenues to explore.
\end{abstract}

%
%



\maketitle

 \section{Introduction}

Natural Language Processing (NLP), despite being a vast domain with numerous dedicated works, continues to present itself as a challenge. The recent proliferation of novel deep neural networks algorithms and methods paved the way to new capabilities in solving NLP tasks ~\cite{Goldberg2016,Ma2016}. 

\begin{figure}[!t]
\includegraphics[width=0.5\textwidth]{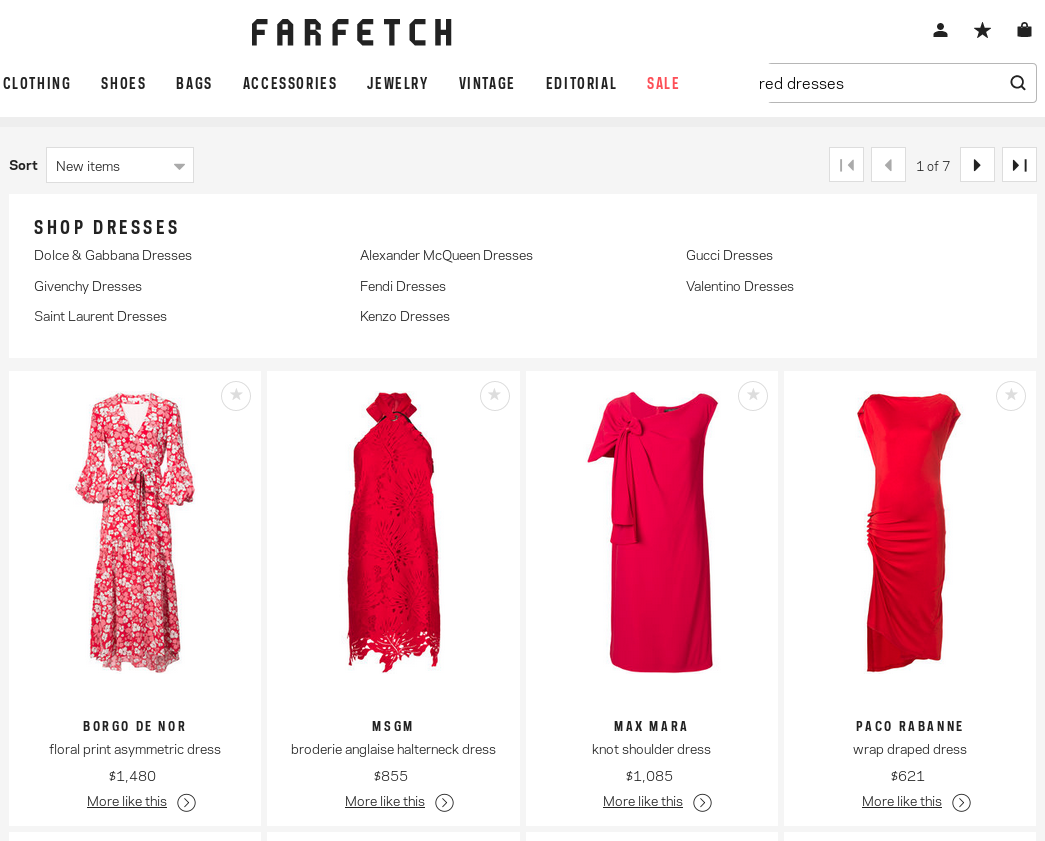}
\caption{Farfetch search engine results for ``red dresses'' queries.}
\label{fig:farfetch}
\end{figure}

At Farfetch, the global platform for luxury, a robust search engine is required to cope with its users' needs and expectations (Figure~\ref{fig:farfetch}). Historically,  the search engine parser has been keyword-based, involving a large number of constraint rules. As Farfetch expands, new ways to provide accurate search answers are required.

Though robust for general purposes, the limitations of the established keyword-based approach are twofold: 1) the inability to capture nuances of fashion epithets (for instance, semantically speaking, what is a floral dress?), and 2) an algorithm able to translate the domain into an effective identification of fashion entities in potentially ambiguous contexts.

\begin{figure*}[!t]
\centering
\includegraphics[scale=0.39]{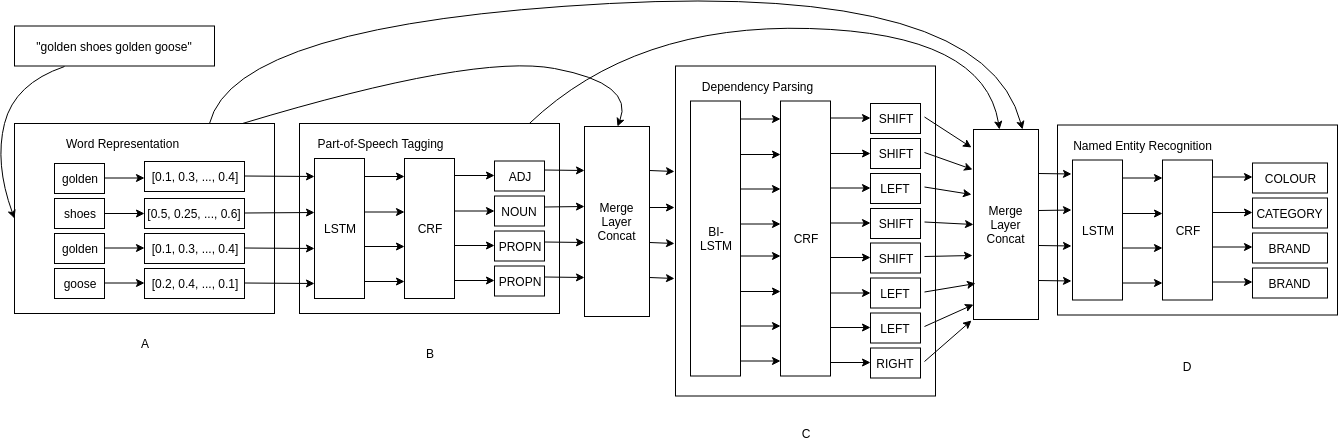}
\caption[Semantic Search architecture diagram for the sentence ``golden shoes golden goose''.]{Semantic Search  architecture diagram for the sentence ``golden shoes golden goose''.}
\label{fig:fullarch}
\end{figure*}

The contributions of this work are the following:

1) we propose a hierarchical deep neural network parser for the fashion domain. Each of the modules of the deep neural network is specialized in a particular task: Word representation, Part-of-Speech tagging (PoS), Dependency Parsing (DP) and Named Entity Recognition (NER) --- see Figure~\ref{fig:fullarch};

2) a transfer learning approach, to cope with the semantics of the fashion domain while reusing information from generic domains. We show the robustness of this network in a fashion dataset for each of the different modules that compose the aforementioned architecture.

This work is structured as follows: In Section~\ref{sec:relatedwork}, we briefly present the current state-of-the-art. 
In Section~\ref{sec:semanticsearch}, our proposal is thoroughly described followed by our preliminary results (Section~\ref{sec:experiments}). Conclusions are drawn in Section~\ref{sec:conclusion}.

\section{Related Work}

\label{sec:relatedwork}

Behind each search engine, there is a drive to present relevant results by taking users' needs and expectations into account. Early works looked at query-document matching algorithm~\cite{Manning2008,Robertson2009} solutions. Although useful, they limit the interpretation and do not take context into account, resulting in models which fail in ambiguity prone environments. Recent advances in the literature lead us to the use of neural networks, providing us with an ability to detect complex non-linear patterns between sets of inputs.  Although not explicitly, it also gives us the ability to leverage together several NLP tasks in a natural way.  Transfer learning domain explores the reuse of pre-trained models and subsequent application to a different set of problems. Inspired by transfer learning, the appearance of holistic architectures to respond to complex NLP tasks~\cite{Collobert2008,Collobert2011} started to emerge, either with many-task solutions \cite{Collobert2011,Hashimoto2016,Chen2016} or simply end-to-end \cite{Chiu2015,Ma2016} architectures.

The quest to a simple unified model, which tries to leverage on multi problems insights in a hierarchical way drives our path. 
The word representation problem explores embeddings techniques in order to map words in a 
vectorial space while being able to capture semantic similarities.  Word embedding solutions presented in literature such as word2vec \cite{Mikolov2013} or GloVe \cite{Pennington2014} establish a starting point towards our objective. 
Although new solutions that explore character embedding techniques emerged, such as fastText \cite{Joulin2016}, we decided to delay their exploration to a later stage. To provide syntactic insights, Part-of-Speech tagging solutions took a leap forward from hidden markov models trends \cite{Brants2000}, with a use of deep end-to-end sequence labelling techniques \cite{Ma2016}. 
The task of associating dependencies between words, Dependency Parsing, evolved into a set of two different approaches. Transition-based architectures, based in Chen \cite{Chen2016} and Dyer \cite{Dyer2015} work, and graph-based architectures \cite{Pei2015} dominate the state-of-the-art solutions to enrich with syntactic association information.

A vastly explored problem, Named Entity Recognition, deals with the identification of entities in textual content. In a similar fashion to the Part-of-Speech tagging task, end-to-end sequence labelling architectures \cite{Chiu2015,Ma2016} dominate recent approach paths. 

In Section~\ref{sec:semanticsearch} we will delve into the details of our proposal, a hierarchical Natural Language Parser.

\section{Semantic Search}
\label{sec:semanticsearch}
Common approaches for information retrieval encompass keyword matching methodologies. Although pragmatic, a caveat of such approaches is that the vast majority rely on rules, hence not being flexible enough or being hard to define.

At first sight, we can identify a few major challenges. Search queries follow a long tail distribution, with several different queries with extremely low frequency. Another problem is related to the semantic gap between queries and documents. The language can also be different depending from which country Farfetch website is accessed. Finally, our vocabulary is also domain specific, making it necessary to develop a tailor-made fashion aware solution.

At this moment, Farfetch’s search is mainly keyword-based and has lots of constraining rules. It lacks the ability to interpret text, often creating noise due to unsolved ambiguities. The Natural Language Parser, which goes by the name of Semantic Search, provides a novel way to interpret text. Given a query, such as \textit{"I want red dresses"}, ideally, the Natural Language Parser should be able to extract the tag \textit{"Colour"} from the word \textit{"red"} and \textit{"Category"} from the word \textit{"dresses"}. We increase the search engine robustness, giving it the ability to interpret more complex and semantically rich queries. Conceptually, the Natural Language Parser is sub-divided in a set of vital steps, namely Word Representation, Part-of-Speech tagging, Dependency Parsing and Named Entity Recognition. Each task has a hierarchical order, forming a pipeline which extracts, at each step, more information contained in the query. 


Semantic Search's Word Representation module starts by slicing each sentence into a sequence of tokens, taking into consideration a text normalization step. We produce an embedding layer on top of the sequence of tokens, which maps words used in similar contexts closer together. In the next step, a Part-of-Speech tagging solution enriches content with sequence labelling methods. This identifies components of a given sentence, such as nouns, adjectives or verbs. Next in line, a Dependency Parser extracts relationships between words and resolves ambiguities concerning inheritance. The final module in the Semantic Search hierarchical pipeline extracts Named Entities, being responsible for identifying key tags (\textit{e.g} Brand, Colour, Category, Materials) which are part of any fashion domain.

\subsection{Word Representation}

The Word Representation module is not only responsible for the word embedding phase but also for the pre-processing concerning normalization.  It starts by slicing each sentence in a sequence of tokens, taking spaces and lower-case capitalization rules into consideration. Although common in the literature, stemming, lemmatization and other more complex operations were not considered. The inherent simplification associated with those techniques would reduce the information content, with potential losses in terms of syntactical context. 

The creation of a dense vector representation can be achieved through several different techniques. In our case we consider GloVe \cite{Pennington2014} and word2vec~\cite{Mikolov2013}. Both are word-based, taking into consideration the frequency of proximal words to optimize their numeric representation. Words that are used in the same context get mapped closer in the vectorial space. 

Each token is associated with an embedded dense vector representation, which are specially prepared to deal with the fashion domain vocabulary, due to its text normalization and training nuances. 
Figure~\ref{fig:fullarch} section [A] illustrates the module architecture. More about the experimental study on Section~\ref{sec:experiments}.

\subsection{Part-of-Speech Tagging}

One of the pivotal tasks in our Natural Language Parser is the identification of components of a given sentence (e.g., noun, adverb, verb). This module is an end-to-end sequence labelling solution, based on a Long-Short-Term-Memory (LSTM) combined with a Conditional Random Fields (CRF) classifier inspired by Xuezhe Ma and Eduard Hovy's \cite{Ma2016} work. It is trained from scratch using an internal dataset, generated from Farfetch products' long descriptions. We argue that, in our platform, search queries are in fact a simplification of products' long descriptions. By learning how to correctly classify long descriptions we are in fact a step closer to correctly classify each query. Figure~\ref{fig:fullarch} section [B] illustrates the module architecture. It takes as input a list of tokens embedded vectors. The next layer is an LSTM with hyperbolic tangent function (tanh) as activation. The recurrent activation is a hard sigmoid. The following layer is a CRF which returns the marginal probabilities on each time step. We use an RMSprop optimizer, a method proposed by Geoffrey Hinton \cite{Hinton2012}. The results and parameters details can be seen on Section~\ref{sec:experiments}.

\begin{table*}[!h]
\small
\centering
\begin{tabular}{l l p{3cm} p{10cm}}
\toprule
\textbf{Brand} & \textbf{Category} & \textbf{Short Description} & \textbf{Long Description} \\ \midrule
 Y / Project & Jeans & High Waisted Jeans with Chaps   & ``Black cotton High Waisted Jeans with Chaps from Y/PROJECT featuring a button and zip fly, belt loops, a five pocket design, a straight leg and frayed edges.'' \\[5pt] 
 Givenchy    & Bags  & Studded Antigona Tote           & ``Dark red calf leather studded Antigona tote from Givenchy featuring a trapeze body, a gold-tone logo plaque, top handles, a gold-tone stud detailing, a main internal compartment, a full lining, an internal zipped pocket, an internal logo patch and a detachable and adjustable shoulder strap.'' \\[5pt] 
 Victoria Victoria Beckham & Shorts & Patterned Shorts & ``Black cotton blend patterned shorts from Victoria Victoria Beckham.''\\[5pt] 
 Fleur Du Mal & Lingerie & Charlotte lace denim thong  & ``Black and indigo blue cotton-blend Charlotte lace denim thong from Fleur du Mal. Underwear and lingerie must be tried on over your own garments.'' \\ \midrule
 
 \toprule
\end{tabular}
\caption{An example of product's short and long descriptions.}
\label{tab:descriptions}
\end{table*}

\subsection{Dependency Parsing}

A syntactic analysis requires not only the identification of a Part-of-Speech tag for each word but also their relation. In this context, a Dependency Parser is responsible for creating a graph connection between each word.

We started by implementing a transition-based dependency parser, inspired by Chen and Manning \cite{Chen2014}. Traditionally, this is a sequential task, in which the input is unfolded into a stack, a buffer and a list of operations. In the initial state, the buffer contains the list of embeddings of the sentence, while the other structures start empty. The stack works as an auxiliary structure, while the list of operations store all the predicted operations so far. During its run, an LSTM unit receives a state representative vector and generates the next operation in line. The next state is computed from the previous one in conjunction with current output. The final state is achieved when both the stack and the buffer get empty. Once reached, we can retrieve the final list of operations.

Although, we were able to mount a solution in a transitional manner, we felt the necessity of implementing a different model, due to the inherent speed problems, as described in Section ~\ref{subsec:experimental_setting}. Thus, inspired by Dyer's \cite{Dyer2015} Stack-LSTM work, we have modified the architecture into an end-to-end sequence labelling problem. Contrary to our previous approach, where the model is run N times until the final state is reached, this new method only requires the recurrent network to run once. For a list of tokens we sequentially predict operations which get appended to the corresponding final list. We explored different architectures and found out a model composed by a Bidirectional-LSTM layer connected to a CRF layer provides more solid results. We also achieve a more robust outcome when using the Part-of-Speech tagging module output as feature. We apply a RMSprop optimizer learning rate method. Figure~\ref{fig:fullarch} section [C] illustrates the module architecture. The results and architecture details can be seen on Section~\ref{sec:experiments}.

\subsection{Named Entity Recognition}

 At last, the Named Entity Recognition module seeks to detect and classify tokens into meaningful tags to Farfetch's domain (e.g., Brand, Colour, Category). Its contributions are key in order to enrich our text interpretation capability.  This module tightly connects our Natural Language Parser to the fashion domain. In terms of technical approach, the task is similar to Part-of-Speech tagging, where each token associates to its respective Named Entity tag. Our end-to-end sequence labelling solution, which explores a similar solution to Xuezhe Ma and Eduard Hovy \cite{Ma2016} work, uses the previously described tasks outcome as features to increase the information flow into itself.

This module is an end-to-end sequence labelling LSTM-CRF, in which each input token vector representation is enriched with its Part-of-Speech tag and Dependency Parsing connections. In practice, we concatenate each token vector with an embedding representation of Part-of-Speech and Dependency Parsing features, token-wise.

Figure~\ref{fig:fullarch} section [D] illustrates the module architecture.
A more detailed explanation and benchmark results are provided on the following Section~\ref{sec:experiments}.

\section{Experimental Study}
\label{sec:experiments}

\subsection{Dataset}
\label{subsec:dataset}

As noted throughout the paper, fashion domain only recently has received significant attention. In this regard, and to the best of our knowledge, the available data on the web for model benchmarking in this domain is scarce. Due to this, our analysis was restricted to Farfetch data. This data comes mainly from three main sources. One of the sources is products' long and short descriptions. Upon each new product arrival, data collecting and generation process starts. In association with a novel product, we have short and long descriptions, consisting of a couple of paragraphs that textually describe the product's characteristics - see Table~\ref{tab:descriptions}. Another source is related with our user's interaction with our search engine. We parse clickstream logs that for each query register its potential association with a clicked product. The last source is related to data augmentation techniques, which leverage existing product characteristics to create artificial content through template-based generation. Those characteristics range from brands to categories (different levels, e.g., $clothing$ $\succ$ $dresses$ $\succ$ \emph{day dresses}), passing through attributes (such as colour) - see Table~\ref{tab:descriptions}.  Afterwards, with data generation techniques associated with template generation we are able to extend our useful vocabulary. We render over 91k unique words and 21.3M annotated sentences.

\subsection{Experimental Setting}
\label{subsec:experimental_setting}

The Natural Language Parser architecture is depicted in Figure~\ref{fig:fullarch}. Each one of the four modules on the architecture is trained separately, in a sequential order. After each module training phase, we freeze its weights and append the next module. We maintain a 9:1 train/test split across all experiments. All reported results are obtained from the testing set.

\smallskip\noindent{{\bf Word Representation:}} 

Firstly, in our embedding layer, we use dense vectors techniques to achieve a semantic word representation. As pre-processing strategy, sentences were split into tokens along white-spaces. Our tokenizer scheme did not take more complex approaches into account, such as stemming or lemmatization, since that would remove important usable information from each sentence. 

The creation of embeddings is a crucial step in our solution. They produce a robust word representation which serves as foundation for each one of the following modules. We only use short and long product descriptions, where each word can be represented by its context. 

We have made an extensive study with two well-known embedding techniques: word2vec\cite{Mikolov2013} and GloVe\cite{Pennington2014}. The hyper-parameters fine tuning was set according to the original papers. We keep an embedding dimension of 300 across all models. 

\smallskip\noindent{{\bf PoS:}}

In the Part-of-Speech tagging task, we aim to learn the correct syntactic annotation for each token. We argue that, in our platform, search queries are a simplification of product long descriptions. By correctly learning annotated long descriptions, the model should easily generalize the Part-of-Speech tags for our search queries.
As training data, we used a pre-annotated dataset of our products' long descriptions. State-of-the-art models, trained with formal English, should be able to generalize their results to our fashion context due to the non-domain specificity of the grammar. Standard models, however, usually rely on language specificities, such as capitalization, for getting higher output confidences. In our use-case, queries are introduced by users, who often neither write perfect sentences nor follow all the referred particularities. Thus, we annotate long descriptions with a state-of-the-art PoS tagging model and apply text normalization on top, so our input data is more representative of what exists in the real world. The objective is to capture the nuances of fashion language constraints in a more limited representation, which do not occur in classical datasets used to benchmark PoS tagging tasks \cite{Marcus1993}. 

Our PoS tagging solution (LSTM-CRF as presented in~\cite{Ma2016}) was trained with 2048 elements per batch, 512 steps per epoch and 32 validation steps. We use dropout layers for regularization. The LSTM layers has 100 units and the CRF is able to classify a total of 20 classes of Part-of-Speech tags, a common simplification from the 36 used in the Penn Treebank Project\cite{Marcus1993}. 

\smallskip\noindent{{\bf DP:}}

The Dependency parsing module takes Part-of-Speech tags and embeddings as features in order to learn the relationship between words, namely how they modify each other. Similarly to the PoS tagging module, we use the long descriptions corpus and their corresponding annotations. Yet again, it is necessary to train a tailor-made solution from scratch.
As stated in Section ~\ref{sec:semanticsearch}, we've implemented two different modules to access this task. We discarded our initial sequential solution, the transition-based iterative module, due to its low training speed. Although the problem is  stated and addressed by \cite{Chen2016}, our data volume severely aggravates this limitation. As previously mentioned, the chosen model consists of an end-to-end architecture using a Bidirectional-LSTM to predict the full list of operations in a single iteration, in a many-to-many fashion.
Comparing to the Part-of-Speech tagging neural network model, besides using the Bidirectional variant of LSTM, we also double the number of units to 200. The CRF layer is able to predict a total of four classes: the three types of operations present in a transition-based solution plus the unknown tag. 
During training we use 2048 elements per batch, 512 steps per epoch and 32 validation steps.
 
 \smallskip\noindent{{\bf NER:}}

Finally, the Named Entity Recognition module is responsible for annotating a range of meaningful domain-specific tags, such as \emph{Brands} and \emph{Categories}. The data produced by each one of the previous models gets concatenated and used as input to the Entity Recognizer, taking full use of the hierarchical construction.
Named Entity Recognition presented some serious challenges, mainly due to the lack of annotated data. Typical solutions do not work in our domain, so the obvious approach is to extrapolate tags from our long descriptions dataset. In order to increase the quality of results, we perform data augmentation via artificial generation as mentioned in Section ~\ref{subsec:dataset}. This way we could reduce length bias and the propagation of the unknown tag across the results. The best model consists of an LSTM of 100 units combined with a CRF layer, similar to \cite{Ma2016} and to our Part-of-Speech tagging implementation. We consider 5 different tags of entities, namely \emph{Brand}, \emph{Category}, \emph{Colour}, \emph{Attribute} and \emph{Unknown}.

\subsection{Results}

\begin{table}[!htp]
\centering

\begin{tabular}{ c l l c c }
\toprule
 \textbf{Task} & \textbf{Model} & \textbf{Features}  & \textbf{Accuracy} & \textbf{F1} \\ \midrule
POS & LSTM+CRF & WORD & 0.9820 & 0.9821  \\
\toprule
DP & BI-LSTM+CRF & WORD+POS & 0.8079 & 0.8043 \\
\toprule
NER & LSTM+CRF & WORD+POS+DP & 0.9902 & 0.9903  \\
NER & LSTM+CRF & WORD+POS & 0.9898 & 0.9898  \\
NER & LSTM+CRF & WORD & 0.9891 &  0.9891 \\

\toprule
\end{tabular}
\caption[Experimental Results]{Experimental Results for all modules in our hierarchical deep learning Natural Language Parser}
\label{restab:all}
\end{table}

Table \ref{restab:all} presents a list of benchmarked models, and their characteristics, to be evaluated against Accuracy and F1 score metrics. 

\begin{table}[h!]
\small
\centering
\begin{tabular}{ c c c}
\toprule

          & \textbf{word2vec}   & \textbf{GloVe} \\ \midrule 
 'skirt'  & ('dress', 0.68)     & ('been', 0.86) \\
          & ('skort', 0.56)     & ('brass', 0.84)  \\
          & ('shorts', 0.55)    & ('short', 0.83) \\
          & ('miniskirt', 0.55) & ('that', 0.81)  \\
          & ('culottes', 0.53)  & ('details', 0.81) \\  \midrule
 'hat'    & ('beanie', 0.73)    & ('loafers', 0.95)  \\
           & ('fedora', 0.68)     & ('thanks', 0.91)   \\
           & ('beret', 0.67)      & ('hinge', 0.91) \\
           & ('trilby', 0.61)     & ('thing', 0.90)  \\
           & ('headband', 0.59)   & ('evokes', 0.90) \\  \midrule
 'Gucci'  & ('fendi', 0.69)     & ('fly', 0.97)  \\
           & ('dsquared2', 0.64)  & ('la', 0.89) \\
           & ('moschino', 0.64)   & ('offers', 0.77)\\
           & ('kenzo', 0.63)      & ('the', 0.75) \\
           & ('givenchy', 0.62)   & ('le', 0.73) \\ 
\toprule
\end{tabular}
\caption[An example of embeddings trained with Farfetch's data.]{An example of embeddings trained with Farfetch's data.}
\label{tab:wordrepresentation}

\end{table}

On Table \ref{tab:wordrepresentation} we can look at some examples of neighbouring words to common fashion vocabulary in our embedding space. We confirm related words seem to get mapped closer to each other.  For example, in the word2vec solution, the word \textit{"hat"} is closer to hat types, such as \textit{"beanie"}, \textit{"fedora"} and \textit{"beret"}. The same happens using the brand name \textit{"Gucci"}, which strongly relates to other brands, such as \textit{"Fendi"} and \textit{"Kenzo"}, in our vectorial space. Due to the unsupervised nature of these tasks, we must perform an extrinsic evaluation oriented to measurable tasks. In our case, we've studied the impact on our models, which behaved similarly for both word2vec and GloVe. We've decided to follow with word2vec in our Word Representation module.

\newcommand{\makecell}[2][@{}c@{}]{\begin{tabular}{#1}#2\end{tabular}}

\begin{table*}[!htp]
\centering
\begin{tabularx}{\linewidth}{c X}
  \toprule
  \textbf{query} & \hfill \textbf{result} \hfill \null \\
  \midrule
  \hspace*{20pt} 'golden shoes golden goose' \hspace*{20pt} &
  \hfill \makecell{\textbf{golden} \\ ('ADJ', 0.85)} 
    \hfill \makecell{\textbf{shoes} \\ ('NOUN', 0.94)}
    \hfill \makecell{\textbf{golden} \\ ('PROPN', 0.80)}
    \hfill \makecell{\textbf{goose} \\ ('PROPN', 0.95)} \hfill \null \\
  \midrule
  'red dress from red valentino' &
  \hfill \makecell{\textbf{red} \\ ('ADJ', 0.99)} 
    \hfill \makecell{\textbf{dress} \\ ('NOUN', 1.00)}
    \hfill \makecell{\textbf{from} \\ ('ADP', 1.00)}
    \hfill \makecell{\textbf{red} \\ ('PROPN', 1.00)}
    \hfill \makecell{\textbf{valentino} \\ ('PROPN', 1.00)} \hfill \null \\
  \bottomrule
\end{tabularx}
\caption{Part-of-Speech tagging.}
\label{restab:pos}
\end{table*}

In more detail, Table~\ref{restab:all} presents the results for the Part-of-Speech tagging task. Alongside the standard LSTM model presented, we also tested its Bidirectional variant. We opted for the simpler architecture due to the lack of value provided by its extra parameters. Table~\ref{restab:pos} displays the results of ambiguous queries to our use case. One kind of challenge our system may encounter is the usage of `golden', either to identify a colour or a brand.  This is where the Part-of-Speech tagging may help reducing ambiguity, since these two forms can only  be distinguish syntactically, either as an adjective or proper noun respectively. A similar problem can be seen for the word `red' (colour or the \emph{Red Valentino} brand).

\begin{figure}[h!]
\centering
\includegraphics[scale=0.5]{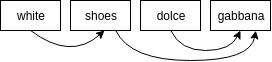}
\caption[Dependency Parsing results for the sentence ``white shoes dolce gabbana'']{Dependency Parsing results for the sentence ``white shoes dolce gabbana''.}
\label{restab:dp}
\end{figure} 
\begin{figure}[h!]
\centering
\includegraphics[scale=0.5]{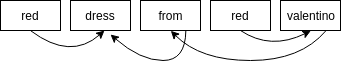}
\caption[Dependency Parsing results for the sentence ``red dress from red valentino'']{Dependency Parsing results for the sentence ``red dress from red valentino''.}
\label{restab:dp2}
\end{figure}

 Again, Table~\ref{restab:all} describes the results associated with our best model for the Dependency Parsing task.
 Contrary to what was registered in the Part-of-Speech tagging solution, the Bidirectional formulation played a crucial role in this task.  In Figure ~\ref{restab:dp} and ~\ref{restab:dp2} we can see some specific results of our model.
 
 \begin{table*}[!htp]
\centering
\begin{tabularx}{\linewidth}{c X}
  \toprule
  \textbf{query} & \hfill \textbf{result} \hfill \null \\
  \midrule
  \hspace*{20pt} 'golden shoes golden goose' \hspace*{20pt} &
  \hfill \makecell{\textbf{golden} \\ ('COLOUR', '0.99')} 
    \hfill \makecell{\textbf{shoes} \\ ('CATEGORY', '0.66')}
    \hfill \makecell{\textbf{golden} \\ ('BRAND', '0.88')}
    \hfill \makecell{\textbf{goose} \\ ('BRAND', '1.00')} \hfill \null \\
  \midrule
  'black gucci bag' &
  \hfill \makecell{\textbf{black} \\ ('COLOUR', 0.63)} 
    \hfill \makecell{\textbf{gucci} \\ ('BRAND', 1.00)}
    \hfill \makecell{\textbf{bag} \\ ('CATEGORY', 0.64)} \hfill \null \\
  \bottomrule
\end{tabularx}
\caption{Named Entity Recognition.}
\label{restab:ner}
\end{table*}
 
 Finally, in Table ~\ref{restab:all} we present the Named Entity Recognition results.  While training we noticed that our models were not able to deal with the dominant 'unknown' tag present in our sparse long descriptions annotations. The template generation was key to increase the results quality.  We achieved best results with a complete architecture using all the information as feature to our model, connected to the LSTM-CRF model. Although the results are extremely similar, those slight variations are the difference between solving ambiguities in a correct way.  Table ~\ref{restab:ner} present some examples queries to that model.

\section{Conclusion}
\label{sec:conclusion}

Despite more than two decades of research and development on search engines, solutions tailored for the fashion domain are still feeble.  Interpretation of textual queries written by users cannot be approached by a keyword-based solution alone.  We provide a solution to comprehend the domain with effective identification of fashion entities.

In this article we have described a hierarchical deep learning Natural Language Parser that takes advantage of transfer learning techniques and specialist models to learn fashion vocabulary and be able to segment and recognize fashion entities in the wild.  Although we focused on the fashion-domain, this approach can  easily be generalized to other contexts. 

Our empirical results establish a robust baseline, which justifies the use of hierarchical architectures of deep learning models. Farfetch now has the ability to interpret text in a novel way enabling the search engine to build much more precise and non-ambiguous queries. Ultimately, this leads to a much better user experience while opening new paths of research to provide tailor-made results.

\begin{acks}
  The authors would like to thank all team members of the search department from Farfetch. Thanks to their support, we could assess the quality of our algorithms and the usability advantages provided to our customers. The first author would also like acknowledge Ernesto Costa from Universidade de Coimbra for supporting his work while at Farfech. 
\end{acks}

\bibliographystyle{ACM-Reference-Format}
\bibliography{bibliography}

\end{document}